# DVS for On-Chip Bus Designs Based on Timing Error Correction

Himanshu Kaul[†]    Dennis Sylvester    David Blaauw    Trevor Mudge    Todd Austin

[†]Circuit Research Labs, Intel Corporation, Hillsboro, OR-97124, USA
Electrical Engineering and Computer Science, University of Michigan, Ann Arbor, MI-48109, USA

**Abstract**

*On-chip buses are typically designed to meet performance constraints at worst-case conditions, including process corner, temperature, IR-drop, and neighboring net switching pattern. This can result in significant performance slack at more typical operating conditions. In this paper, we propose a dynamic voltage scaling (DVS) technique for buses, based on a double sampling latch which can detect and correct for delay errors without the need for retransmission. The proposed approach recovers the available slack at non-worst-case operating points through more aggressive voltage scaling and tracks changing conditions by monitoring the error recovery rate. Voltage margins needed in traditional designs to accommodate worst-case performance conditions are therefore eliminated, resulting in a significant improvement in energy efficiency. The approach was implemented for a 6mm memory read bus operating at 1.5GHz (0.13μm technology node) and was simulated for a number of benchmark programs. Even at the worst-case process and environment conditions, energy gains of up to 17% are achieved, with error recovery rates under 2.3%. At more typical process and environment conditions, energy gains range from 35% to 45%, with a performance degradation under 2%. An analysis of optimum interconnect architectures for maximizing energy gains with this approach shows that the proposed approach performs well with technology scaling.*

## 1. Introduction

On-chip buses can contribute a significant portion of the total power consumption. This is especially true for high performance and communication-centric designs, where the buses are long and are heavily buffered to meet aggressive delay targets. Due to increased wire delays and longer wire lengths, on-chip communication is often performance critical and directly impacts the processor cycle time. The design parameters of a bus (pitch, number and size of repeaters, shield wires, etc.) must be chosen to meet the timing constraints under worst-case conditions, including switching behavior of neighboring wires, process corner and environment (IR-drop and temperature) conditions. The probability of all worst-case conditions occurring simultaneously is usually small and hence, the bus is faster than it needs to be for more common case operating conditions. In this paper, we propose a method for dynamically scaling down the supply voltage for typical case conditions, resulting in significant energy reduction while still meeting delay constraints.

Various layout [1,2], repeater sizing [3,4] and encoding [5,6] solutions have been proposed for reducing power consumption in on-chip buses. However, these approaches are focused on improving energy efficiency at the worst-case conditions and do not take advantage of the potential energy reduction at more typical operating conditions. Hence, they are orthogonal to the approach proposed in this paper. Also, supply voltage scaling is a more powerful method for energy reduction, since it can ideally result in a quadratic savings in energy, without any routing area overhead. Dynamic voltage scaling (DVS) for energy reduction is not a new concept in general, but its application for on-chip buses operating in non-worst-case conditions has not been explored previously.

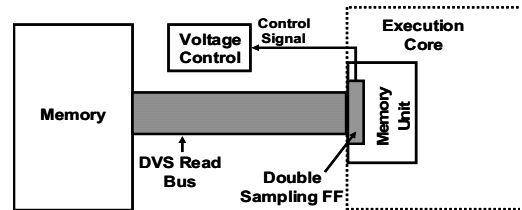

**Fig. 1. DVS memory read bus with double sampling flops.**

The traditional approach to DVS exploits fluctuations in the computational requirement of an application and scales down the frequency and voltage during periods of low performance utilization [7,8]. However, this approach does not exploit the variation in operating conditions and therefore still incorporates substantial margins to accommodate worst-case conditions. In this paper, we formulate the problem as reducing energy while maintaining the same clock frequency, noting that this approach can be combined with traditional DVS as well. A number of approaches were proposed that embed self-tuning circuits to provide dynamic reduction of the supply voltage and/or clock frequency.

The so called *Correlating VCO* [9,10] and *Delay Line Speed Detector* [11] schemes use a circuit under test that mimics the critical path delay on the chip and the voltage for the chip is changed based on the speed of this test circuit. Duplicating a bus would result in a huge area overhead, making such techniques impractical for designs with buses as the critical path. In a different tuning approach, the actual circuit performance is tested periodically with worst-case latency vectors using a so-called *Triple-Latch Monitor* [12]. Though this approach can adjust to more local performance conditions, it cannot take advantage of typical latency vectors. Also the power overhead can be substantial for bus designs since worst-case vectors need to be propagated through the bus for evaluating the operating condition at regular intervals. Finally, IR-drop at repeater blocks in a bus are strongly dependent on the input vectors due to the large size of repeaters and their influence on IR-drop. In this case, the *Triple Latch Monitor* cannot take advantage of typical delays and IR-drop on a bus.

A key characteristic of the previous approaches is that they ensure correct operation at all times, and hence require additional safety margins which reduce their power efficiency. This paper proposes a more aggressive voltage scaling technique that is based on dynamic detection and correction of delay errors. Error recovery is incorporated in the bus architecture by employing a modified flip-flop that samples its input at the normal clock as well as at a delayed clock and was previously proposed for use in logic pipeline designs in [13]. If a difference between the 2 samples is detected, a control signal from the flop indicates that the data captured by the normal clock is incorrect and an error recovery mechanism ensures that the correct data value (that was sampled by the delayed clock) is propagated, while also ensuring that the incorrect data from the previous cycle is flushed out from the next stage.

A major advantage with the proposed approach is that error recovery does not require a failing vector to be retransmitted on the bus, providing great potential for energy reduction on the bus with a smaller energy overhead for error recovery. At the same time, all local performance conditions, including IR-drop and neighboring switching patterns, are accounted for by the approach, allowing for the removal of all voltage safety margins. In the proposed approach, the voltage is increased only when not doing so would result in an unacceptable number of delay error corrections, thereby significantly improving the energy efficiency.

We implemented the proposed approach for a memory read bus of an Alpha processor design. Fig. 1 shows the schematic of the system level implementation that would be required for this approach. The bus feeds into the memory unit of the execution core, where load data is typically held in a buffer before being committed to an architectural state. The original flip-flops that hold the load data can be replaced by the double-sampling flips-flops and timing errors can be handled in a manner similar to cache misses and speculative loads, with a one cycle penalty for error recovery. A similar mechanism will be required for handling the non-deterministic latency (in cycles) with this approach for other types of buses.

The remainder of this paper is organized as follows. The delay error detection and correction flip-flop is discussed in Section 2, along with its implications on the bus design. The simulation framework is discussed in Section 3. The impact of statically scaled supply voltage on error-rates and energy reductions for a range of benchmark programs is presented across a range of possible process, voltage drop and temperature (PVT) corners in Section 4. In Section 5, a simple voltage regulation system in conjunction with the proposed bus design is simulated for gauging the energy savings possible by dynamically scaling supply voltages to take advantage of the difference in switching activities across programs. Interconnect architectures that favor increased voltage scaling for a bus with the proposed approach are discussed and quantitatively analyzed in Section 6. The implication of this is discussed in the context of the efficacy of the proposed bus design for scaled technologies. Section 7 presents some concluding remarks.

## 2. Double sampling flip-flop Design

The proposed approach to DVS is based on the usage of a double sampling flip-flop (Fig. 2) for error detection and correction. In the absence of timing errors, the flop operates as a traditional master-slave flip-flop. When the input of the flop does not meet the setup time for error-free operation, the correct input is captured by the shadow latch since it is controlled by a delayed clock (with respect to the main flip-flop). The presence of a timing error in the previous cycle is signaled by the *Error_L* signal, which is an XOR function of the data in the slave latch and the shadow latch. Upon assertion of the *Error_L* signal, the correct data is restored in the flop through the multiplexer placed in the feed back path of the master latch. The design approach does need to be conservative to allow for data to be captured correctly by the shadow latch at the minimum allowable voltage under worst-case operating conditions. The local error signals (*Error_L*) of all the individual flip-flops in a bank that lie between two pipeline stages are ORed to produce an error signal that indicates a timing error in the previous pipeline stage. This signal is polled by the control system to measure error rates and is also used for triggering the error recovery mechanism in the architecture. Error correction requires at least a one cycle penalty since the incorrect data that was sent to the next stage needs to be flushed out before the correct data from the shadow latch is re-transmitted. The extra cycle penalty for timing errors could be accommodated by the processor in similar ways that cache misses are handled and the performance (IPC) may not necessarily degrade by the same amount as the error-rate (especially for out-of-order execution).

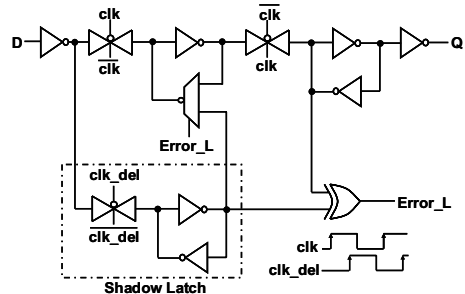

**Fig. 2. Circuit schematic of double sampling flip-flop.**

This error detection and correction capability comes at the cost of a much increased hold-time constraint with the flip-flop. As a result it needs to be ensured that the delays of short paths that feed into a shadow latch never violate the increased hold-time constraint. This hold constraint limits the amount of clock delay that can be accommodated on the shadow latch and hence the degree of voltage scaling below the point of first failure that can be exploited in the DVS scheme. In bus structures, however, the difference between short and long path delays is much less than that in logic circuits, making bus structures highly suitable for this approach of error detection and correction. In our analysis, it was found that the shadow latch clock could be delayed by as much as 33% of the clock cycle without violating the short-path constraint.

With each error recovery in the flop, a performance and energy penalty is incurred. The energy penalty with error-recovery stems from the energy consumed in the error detection/recovery logic and the re-execution of instructions when recovering from an error. As supply voltage is lowered the energy savings increase at the cost of increased error-rates. The trade-off between bus energy reduction and the energy penalty associated with increasing error rates as supply voltage is scaled down implies an optimal supply voltage exists at which the total energy with the DVS approach is optimal. Hence, the voltage must be controlled such that the energy optimal error rate is not exceeded. In addition, the performance impact of error recovery places limits on the acceptable error rate. In our experiments, a maximum average error rate of 2% was used, and was found to provide substantial energy savings while incurring negligible performance impact.

## 3. Simulation framework

A 6mm 32-bit bus (Fig. 3) routed on a global metal layer of a 0.13μm CMOS process at minimum pitch (0.8μm) is used for analyzing the impact of the DVS approach on a memory read bus. A nominal supply of 1.2V is used. Capacitance extraction is performed with a 2D field-solver. A 1.5mm inter-repeater distance is used with shield wires inserted after every 4 wires. Such a shield insertion interval (in terms of wires) is a typical design practice for limiting noise and inductive effects for wide buses [14]. The receiver end of the bus feeds into the input of a flip-flop (not shown in the figure). A fixed clock frequency of 1.5GHz is assumed. The repeaters are sized so that the maximum delay (measured from node *in* to node *out*) on the bus is 600ps (allowing 10% cycle time slack for set-up time and clock skew). The maximum delay is measured under worst-case conditions of neighbor switching activity and the PVT conditions, i.e., slow process corner, temperature of 100C and a voltage (IR) drop of 10%. This sizing approach reflects a typical design philosophy to ensure that the performance target is achieved even under worst-case conditions with a fixed supply voltage.

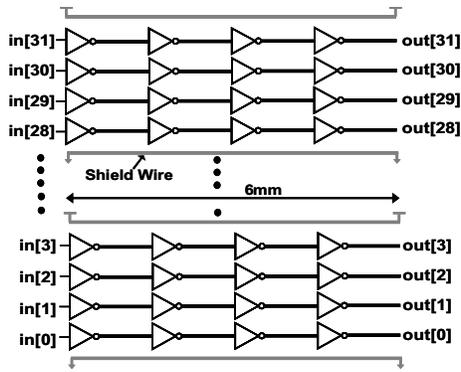
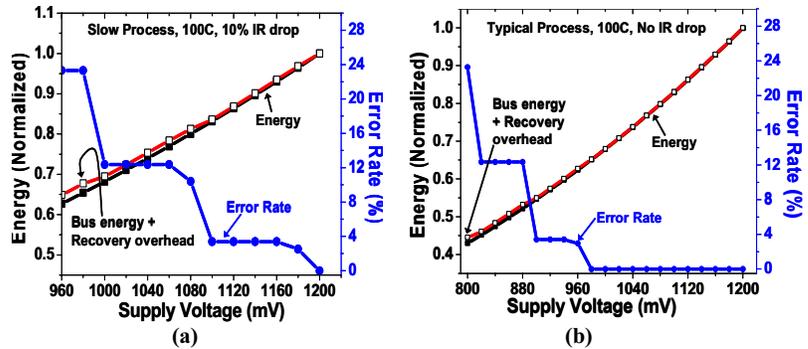

**Fig. 3. Bus setup.**

**Fig. 4.** Energy and error rate analysis for scaled supply voltages for PVT corners of (a) slow process, 100C, 10% IR drop and (b) typical process, 100C, no IR drop.

The bus is analyzed for performance and energy with the DVS approach over millions of cycles of program execution. In order to reduce the simulation complexity, while maintaining SPICE-level accuracy, the delays (for every wire) and energy consumption on the bus are tabulated for all possible data input combinations using HSPICE. Such look-up tables are created for individual supply voltages (in increments of 20mV) over a range of supply voltages and also for different combinations of process corner and temperature. Leakage current through the repeaters is also tabulated for the different supply voltages and environment conditions so as to include the contribution of leakage energy to the total bus energy.

For evaluation purposes we use the data trace on the memory read bus from 10 of the SPEC2000 benchmarks. The data trace for each benchmark is obtained by modifying the sim-safe simulator in the SimpleScalar/Alpha version 3.0 toolset [15]. The simulations were run using the SPEC reference inputs. We used the SimPoint toolset's Early SimPoints to pinpoint 10 million instruction trace sequences that were highly representative of the entire program execution [16]. Each instruction is assumed to represent a single clock cycle (IPC=1). All error-rate analysis is based on the number of resulting bus timing errors (with correct data captured by the shadow latch) during a window of program instructions. A single bus timing error represents the assertion of the error signal by one or more error detecting flip-flops in the bank in a single cycle. The number of errors based on clock cycles for the actual system architecture will yield lower error-rates since the same number of errors will occur in a larger time window due to the fact that IPC in a pipeline is typically less than 1. Therefore, the error rates reported in the paper are pessimistic. From a performance perspective, the reduction in IPC for a particular error rate is highly dependent on the architecture and the specific program. For the sake of simplicity, we assume a 1 cycle penalty for error recovery and translate this to a reduction in performance (IPC) that is the same as the error-rate.

## 4. Voltage scaling and PVT corner impact on error rate

Since the proposed approach relies on error correction at aggressively scaled supply voltages, it is necessary to gauge the effect of scaling supply voltages under different PVT corners on the resulting error rates since it can determine the performance degradation and the energy overhead from error correction. The different process corners used are slow, typical and fast. For local voltage conditions on the repeaters/drivers, either no IR drop is assumed or a 10% droop in supply voltage is assumed for any particular supply voltage when delay is calculated. The temperature conditions assumed are either 25C or 100C. For different combinations of process, IR drop and temperature, the benchmark programs were simulated over a range of supply voltages for all PVT corners.

The effect of scaled supply voltages for 2 different PVT corners is shown in Fig. 4. Every point on the energy and error-rate curves represents combined energy and error rates of running all the benchmark programs (with each one being run for 10 million cycles) at the specific supply voltage. The supply voltage is scaled only up to the point where the longest bus delay can still meet the setup time of the shadow latch for the specific PVT corner. Since the bus was designed to operate error-free for the worst case condition (same as that used for Fig. 4a), the error rates increase as soon as the supply voltage is lowered below the nominal 1.2V supply. For a faster PVT corner, the same performance can be maintained while supply voltage can be scaled. This is evident in Fig. 4b, where no errors are introduced up to a 980mV supply.

For every error, there is an energy overhead involved in re-transmitting the correct data to the processor pipeline. Since only a small fraction of the flops in a bank typically result in errors, most of the extra energy consumption usually comes from clocking all the flip-flops for an extra cycle. The effect of this over head is also shown in Fig. 4 and is very small compared to the energy savings on the bus. The error-correction overhead will be higher when the entire pipeline is considered and would also depend on the architecture. Since we are examining the bus in isolation, we choose error rates from the perspective of performance degradation, which closely tracks the error-rates.

The effect of 3 different target error rates (0%, 2% and 5%) is examined. If a PVT corner results in a faster bus, the supply voltage can be scaled further to achieve the same error rate under the delay constraint. In Fig. 5 the delay spread on the normal bus (at VDD=1.2V) from the various PVT corners is shown on the X-axis. The energy gains achievable by operating the bus at the lowest supply voltage that does not result in exceeding the target error rate for the PVT corner are shown on the Y-axis. As expected, the energy gains for a target error rate increase as the PVT corner results in a faster bus, with gains of 35% for the typical process corner with no performance degradation. Also, if the higher error rate of 5% can be tolerated, the gains are higher. The gains from 0% and 2% error rates are indistinguishable. This results from the fact that the error rates jump directly from 0 to above 2% for the used supply voltage discretizations (of 20mV).

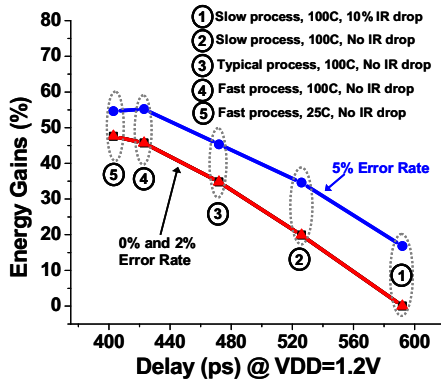

**Fig. 5. Energy gains for target error rates over the delay spread (for non-DVS bus) that can occur from the range of PVT corners.**

Though we treat each of the performance dictating factors independently (for the sake of simplicity), the process corner is the only true independent variable while the IR drop and temperature can also be functions of the specific program. Incorporating such dependencies would involve complex models. The goal of the analysis is to show the range of energy gains possible for the range of performance dictating conditions.

## 5. Energy Reduction with Proposed DVS Scheme

The bus switching activity can vary from one program to another and even during a single program execution. The results of the previous section do not reflect the energy gains possible if the supply voltage is scaled during program execution while still maintaining a target error rate. The in-situ error rate measurement capability of the proposed approach can allow further energy gains (on top of those available from PVT corners) for individual programs by taking advantage of the bus switching activity.

To illustrate the dependence of dynamic supply scaling on program behavior we first examine the optimal supply voltage selection (with the knowledge of future program switching behavior) over time while maintaining a fixed error rate. For different target error rates, the percentage of time that the bus spends over various supply voltages during the execution of 3 different programs is shown in Fig. 6. When no errors are tolerated (not shown in figure), all the programs run at the supply voltage that the specific PVT corner allows for zero error operation - the energy gains being dictated only by the PVT corner. If a small percentage of error rates can be tolerated, the optimal supply voltages during program execution can vary widely from one program to another. For a target error rate of 2% *crafty* can run at 900mV during most of the program execution, while the supply cannot be reduced below 980mV even with a target error rate of 5% for *mgrid*.

In an actual system, it is not possible to guarantee a target error rate since there is delay involved in changing the supply voltage

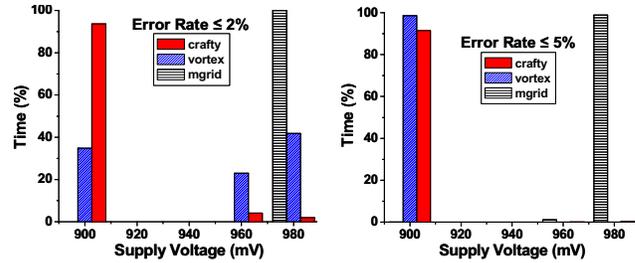

**Fig. 6. Optimum supply voltage distribution during execution for 3 programs while maintaining constant error rates. The PVT corner used is typical process, 100C and no IR drop.**

with a regulator and the switching activity for a block of time in the future cannot be known a priori. A simple voltage supply control system (Fig. 7) is simulated with the bus during program execution. The system calculates errors generated by the bus in 10,000 cycles using an error counter that is incremented by the *Error* signal from the bank of flip-flops. The counter is reset after every 10,000 cycles. The voltage controller changes the voltage based on the error rate in the last 10,000 cycles. The system tries to maintain error rates between 1% and 2% (which is a reasonable trade-off of performance for energy gains). If the error rate is less than 1%, the supply voltage is reduced by 20mV and if it is greater than 2%, supply voltage is increased by 20mV. A more sophisticated proportional control system could have been used that results in voltage changes proportional to the magnitude of error difference between the target and sampled error rates. Since the error-rate of the bus is a non-linear function of supply voltage, it is not possible to calculate the transfer function for the bus. Calculation of the proportionality constant for such a system would not be trivial. Also, the simpler system that we have simulated is shown to work reasonably well without the hardware overhead of a more sophisticated system. Since the voltage regulators take time to adjust the voltage (typically around 1μs/10mV), the supply voltage on the bus is changed by 20mV only after a delay of 2μs (3000 cycles at 1.5Ghz operation) after the decision to change the supply voltage is taken by the regulator. The minimum voltage allowed by the regulator is chosen conservatively for the bus to meet the setup time of the shadow latch. The only factor that we have used for tuning (the minimum allowable voltage by the regulator) is the process corner since that does not change with time. Otherwise, worst case conditions of temperature and IR drop are assumed. For example, if the PVT corner is typical, 100C and no IR drop, minimum voltage allowed

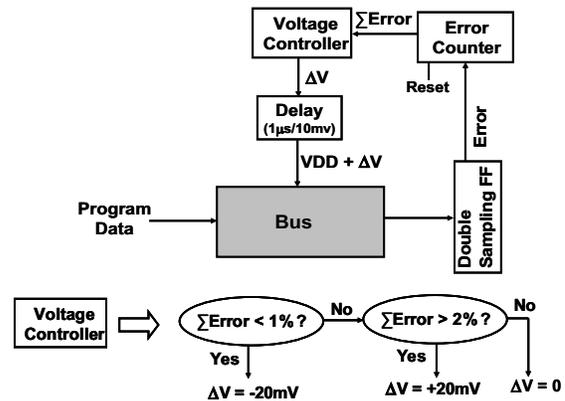

**Fig. 7. Dynamic control system for supply voltage scaling.**

Table 1. Energy gains with 2 DVS schemes for 2 PVT corners.

| Benchmark | Slow Process, 100C, 10% IR drop | | | Typical Process, 100C, No IR drop | | |
|---|---|---|---|---|---|---|
| | Fixed VS | Proposed DVS | | Fixed VS | Proposed DVS | |
| | Energy Gains (%) for 0% error rate | Energy Gain (%) | Average Error Rate (%) | Energy Gains (%) for 0% error rate | Energy Gain (%) | Average Error Rate (%) |
| 1. crafty | 0.0 | 15.4 | 1.62 | 17.0 | 44.6 | 1.15 |
| 2. vortex | 0.0 | 3.4 | 1.53 | 17.0 | 36.9 | 1.53 |
| 3. mgrid | 0.0 | 1.2 | 1.98 | 17.0 | 34.8 | 1.86 |
| 4. swim | 0.0 | 1.2 | 2.22 | 17.0 | 34.6 | 1.94 |
| 5. mcf | 0.0 | 16.4 | 2.23 | 17.0 | 44.7 | 0.91 |
| 6. mesa | 0.0 | 17.5 | 1.89 | 17.0 | 45.2 | 0.95 |
| 7. vpr | 0.0 | 3.0 | 1.51 | 17.0 | 36.6 | 1.52 |
| 8. applu | 0.0 | 2.1 | 1.68 | 17.0 | 35.7 | 1.33 |
| 9. gap | 0.0 | 12.8 | 1.52 | 17.0 | 43.2 | 1.37 |
| 10. wupwise | 0.0 | 1.7 | 1.48 | 17.0 | 35.6 | 1.34 |
| Total | 0.0 | 6.3 | 1.77 | 17.0 | 38.6 | 1.4 |

is such that shadow latch setup time is met for typical process, 100C and 10% IR drop.

Table 1 lists the energy reductions possible with two voltage scaling schemes. The fixed voltage scaling (VS) scheme is a generic representation of other voltage scaling schemes that can only account for global process variations and have conservativeness built into them since they cannot handle timing errors. The proposed DVS scheme is as discussed earlier in the section. For the worst-case PVT corner (slow process, 100C, 10% IR drop), no energy gains are possible with zero error rates (fixed VS), while the implemented DVS results in 1% to 17% energy gains across the benchmarks. Any energy gains at this PVT corner with the proposed DVS are possible only by taking advantage of the unique program switching activities on the bus. Though the combined error-rate for all the programs with the implemented DVS system is less than 2%, 2 programs result in average error rates that are slightly higher than the target. Note however, that while the error rate cannot be guaranteed, correct operation using error recovery is always ensured. At the faster PVT corner (typical process, 100C, no IR drop), the difference between fixed VS and implemented DVS is even higher because the fixed VS scheme can lower supply voltage only up to the point where error-free operation is guaranteed with 10% VDD drop. The fixed VS scheme cannot take advantage of the fact that there may not be any VDD drops in the actual design. At the faster PVT corner all programs finish with individual average error rates within 2% and overall energy gains of 38.6%, while individual programs show gains from 35% to 45%.

The variation in supply voltage and instantaneous (i.e., over a time period of 10000 cycles) error rates for the faster PVT corner of Table 1 are plotted in Fig. 8. The supply voltage is assumed to start from the nominal 1.2V. The programs are run consecutively (each one running for 10 million cycles) and the region of individual program executions is demarcated by the same numerical ordering as in Table 1. For the error rates, each dot represents the error-rate over the period of 10000 cycles. It is evident that the proposed DVS control system can adjust for the PVT corner as well as the switching activity that is unique to every program. The supply voltage and error-rates exhibit unique patterns for each program and the switch from one program to another is clear. As mentioned above, though the average error-rates (from the execution of 10 million instructions of a program) for the programs are within the 2% target, the instantaneous error-rates can be well over the target (reaching as high as 6% in Fig. 8) during execution. The main reason for this is the delay involved in ramping up the voltage by the controller when high error rates are observed.

## 6. Interconnect Architecture and Technology Scaling

The proposed approach to DVS provides advantages similar to that of asynchronous design – maximize gains for typical case operation rather than worst-case. The maximum amount of supply lowering with the proposed approach is primarily limited by the error rates. For a particular error-rate constraint, a design that does not change the worst-case delay while improving the typical case delay allows the supply voltage to be lowered further before the error rate constraint is reached.

The electrical equivalent of an interconnect wire is shown in Fig. 9. The line of interest has a signal $V$ while $A0$ and $A1$ serve as neighboring aggressors. Switching pattern I results in worst-case delays on the wire. The Elmore delay for such a switching pattern is :
$$t_D = R \cdot (C_g + 4 C_c) \quad (1)$$
This switching pattern is responsible for the initial (and acceptable) increase in error rates as supply voltages are reduced. When the supply voltage is reduced to a point where delay (for error-free operation) cannot be met with switching pattern II, the error rates immediately jump to a higher, and often unacceptable level (>10%). The difference in Elmore delays between pattern I and pattern II is: $\Delta t_D = R \cdot C_c \quad (2)$

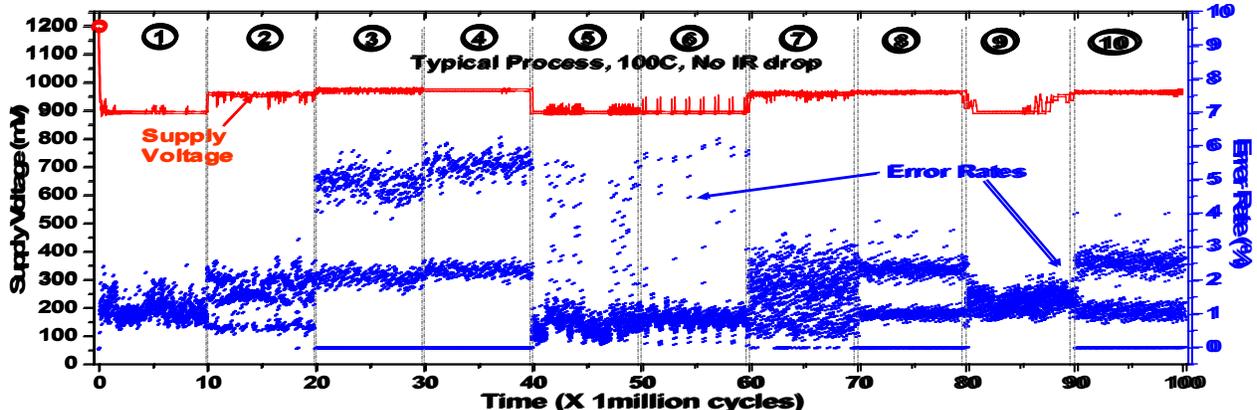

Fig. 8. Supply voltage and instantaneous error rates during execution of the benchmark programs for the PVT corner of Typical Process, 100C, No IR drop. Regions of benchmark execution are labeled by the same numbers as the numerical order in Table 1. Error rates may appear overlapped in time due to resolution limitations.

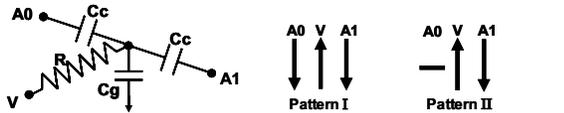

Fig. 9. Electrical equivalent of an interconnect line and two neighbor switching patterns for a victim line.

Wire layout geometries that increase this difference between these delays (by increasing the Cc/Cg ratio) allow the bus to operate at lower supply voltages with the proposed DVS method for the same non-zero error rate constraint, provided the worst case delay remains unchanged. Since the fastest delay on the bus reduces with this approach, the delay to the shadow latch clock would need to be reduced. This limits the minimum voltage that the bus can operate at. Since the power savings with this approach are limited more by the acceptable error-rate rather than the absolute minimum supply voltage that can be operated, this is a reasonable trade-off.

We alter the wire parasitics of the bus so that the Cc/Cg ratio is 1.95X that of the original bus while ensuring that the wire resistance and total effective capacitance (Cg + 4Cc) for worst-case delay does not change. Repeater sizes are unchanged since the worst-case delay does not change. The worst-case switching delay remains unchanged across all PVT corners. Static voltage scaling at various PVT corners confirmed that the voltage could be scaled by a further 20mV (than original bus) for almost all PVT corners before reaching the error-rate constraint. The results for energy gains across PVT corners are shown in Fig. 10. The curve for zero error rates does not change since the maximum delay does not change. The 2% and 5% error-rate curves show slightly higher energy gains. Simulations with the proposed voltage control system of the previous section also showed increased energy gains for all programs, with the average energy gain for the worst-case PVT corner (slow process, 100C, 10%IR drop) increasing from 6.3% (in Table 1) to 8.2%. The error rate for all the programs with the proposed DVS bus is still within the target of 2%.

With scaled technologies, the wire capacitance does not change appreciably [17], while the wire resistance increases. As a result, the delay spread on wires due to neighbor switching activity increases (since the R • Cc term in (2) increases). The proposed bus design results in a higher energy savings with an increased difference in delay between worst-case and more typical switching activities and, therefore, can be expected to scale well with technology.

## 7. Conclusions

A DVS approach for energy reduction in on-chip buses with double sampling flip-flops has been proposed. The proposed bus design allows aggressive voltage scaling since it provides recovery from timing errors without retransmission of failed data on a bus. A 6mm memory read bus that was designed to operate at 1.5Ghz (for a 0.13μm CMOS technology) under worst-case conditions was analyzed using this approach. An extensive analysis of the effect of static voltage scaling across a range of PVT corners showed that if voltage is scaled down to reduce any available slack (while maintaining the same clock frequency) at a particular PVT corner, energy gains up to 48% are possible even with no error rates. The gains increase if slightly higher error-rates can be tolerated. A voltage regulation system to allow dynamic voltage scaling and take advantage of typical switching activities during program execution was also tested. Even at the worst-case process and environment conditions, energy gains up to 17% were achieved for individual programs, while at more typical process and environment conditions, the energy gains range from 35% to 45%. These energy gains were achieved with less than 2% impact on performance. The effect of interconnect geometries that favor increased voltage scaling for a bus with the proposed approach was analyzed and the results indicate that the approach should scale favorably with interconnect scaling.

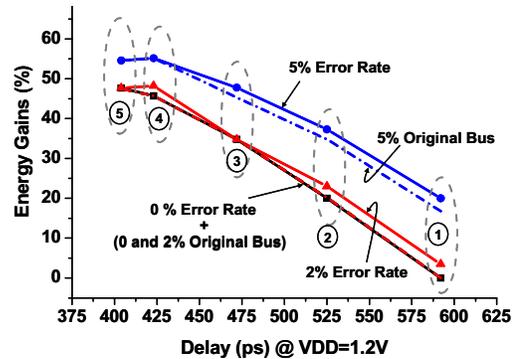

Fig. 10. Energy gains for target error rates over the delay spread (for non-DVS bus) from the range of PVT corners for the modified bus. The PVT corners used are the same as Fig. 5.